# Thermodynamics of amide + ketone mixtures. 1. Volumetric, speed of sound and refractive index data for *N,N*-dimethylformamide + 2-alkanone systems at several temperatures

ANA COBOS[(1)], FERNANDO HEVIA[(1)], JUAN ANTONIO GONZÁLEZ*[(1)], ISAÍAS GARCÍA DE LA FUENTE[(1)], AND CRISTINA ALONSO TRISTÁN[(2)]

[(1)] G.E.T.E.F., Departamento de Física Aplicada, Facultad de Ciencias, Universidad de Valladolid, Paseo de Belén, 7, 47011 Valladolid, Spain,

[(2)] Dpto. Ingeniería Electromecánica. Escuela Politécnica Superior. Universidad de Burgos, Avda. Cantabria s/n. 09006 Burgos, Spain

*e-mail: jagl@termo.uva.es; Fax: +34-983-423136; Tel: +34-983-423757

**Abstract**

Densities, $\rho$, speeds of sound, $c$, and refractive indices, $n_D$, have been measured for the systems *N,N*-dimethylformamide (DMF) + propanone, + 2-butanone, or + 2-pentanone in the temperature range from 293.15 to 303.15 K and at 298.15 K for the DMF + 2-heptanone mixture. Due to the high volatility of acetone, the corresponding $n_D$ measurements were developed at 293.15 K and 298.15 K. The direct experimental data were used to determine the excess molar volumes, $V_m^E$, and the excess refractive indices, $n_D^E$, at the working temperatures. Values of the excess functions at 298.15 K, for the speed of sound, $c^E$, the isentropic compressibility, $\kappa_S^E$ and for the excess thermal expansion coefficient, $\alpha_p^E$, were also calculated. The investigated systems are characterized by strong amide-ketone interactions, which become weaker when the alkanone size is increased. This is supported by negative $V_m^E$ values; by the dependence on temperature and pressure of $V_m^E$, and by positive $P_{int}^E$ (excess internal pressure) values. Analysis of the systems in terms of the Rao's constant indicates that there is no complex formation. In addition, negative $V_m^E$ values also reveal the existence of structural effects, which largely contribute to the excess molar enthalpy, $H_m^E$. $V_m^E$ and $H_m^E$ values increase with the chain length of the 2-alkanone. It allows to conclude that the relative $V_m^E$ variation with the ketone size is closely related to that of the interactional contribution to this excess function. Molar refraction values, $R_m$, show that dispersive interactions become more relevant for the systems including longer 2-alkanones.

1. **Introduction**

Amides are interesting compounds as they are very polar solvents which can form hydrogen bonds both through the oxygen and nitrogen atoms. Particularly, *N,N*-dimethylformamide (DMF) is an aprotic protophilic solvent, with good donor-acceptor properties, capable to solve many organic substances. DMF has many practical applications. It is used in the production of acrylic fibers, plastics, pesticides or surface coatings [1], or as a highly selective extractant for the recovery of aromatic and saturated hydrocarbons from petroleum feedstocks [2]. DMF is also very effective in nanotechnology [3-5]. On the other hand, the investigation of liquid mixtures containing the amide functional group is needed for a better understanding of complex molecules of biological interest [6]. In this framework, DMF is useful as a model compound for peptides, and the aqueous solution of DMF is a simple biochemical model of biological aqueous solutions [7,8]. From a theoretical point of view, amides are also a very interesting class of compounds, as in pure state, they show a significant local order [9]. In the case of *N,N*-dialkylamides, due to the lack of hydrogen-bonds, this has been attributed to the existence of strong dipolar interactions [10]. 2-Alkanones ($CH_3CO(CH_2)_{u-1}CH_3$) are polar, aprotic compounds and hydrogen bonds acceptors, used as solvents for plastics and some synthetic fibers. Acetone is used as intermediate of methyl methacrylate. Ketones are also important substances in Biochemistry, as many sugars are ketones (fructose, e.g.) and fatty acid synthesis proceeds via these compounds. On the basis of these features, the investigation of amide + ketone mixtures seems to be convenient. In this first article of the series, we pay attention to DMF + ($CH_3CO(CH_2)_{u-1}CH_3$) ($u = 1,2,3,5$) systems, reporting density, $\rho$, data, speeds of sound, $c$, and refractive indices, $n_D$, in the temperature range from 293.15 to 303.15 K for systems with $u = 1,2,3$ and only at 298.15 K for the mixture with $u = 5$ since the behaviour of this system is nearly ideal (see below). In the case of the propanone system, $n_D$ measurements were conducted at 293.15 K and 298.15 K due to the high volatility of the compound. A literature survey shows that there are available density data at 298.15 K for systems with propanone or 2-butanone; excess molar volumes, $V_m^E$, for mixtures at 303.15 K including 2-butanone or 2-pentanone [11-13], and excess molar enthalpies, $H_m^E$, for mixtures at 298.15 K with $u = 1,2,3$ [14,15].

2. **Experimental**

*Materials.* Information on the source and purity of the chemicals used is given in Table 1. The compounds were used without further purification. The physical properties, $\rho, c, n_D$, thermal expansion coefficient, $\alpha_p$, and adiabatic, $\kappa_S$, and isothermal $\kappa_T$, compressibilities of the pure liquids are collected in Table 2. Our values are in good agreement with those from the literature.

*Apparatus and procedure.*

Binary mixtures were prepared by mass in small vessels of about 10 cm$^3$, using an analytical balance HR-202 (weighing accuracy ± 0.01 g), with all weighings corrected for buoyancy effects. The error on the final mole fraction is estimated to be lower than ± 0.0008. Molar quantities were calculated using the relative atomic mass table of 2011 issued by IUPAC [16].

Densities and speeds of sound of both pure liquids and of the mixtures were measured by means of a vibrating-tube densimeter and sound analyser, Anton Paar model DSA-5000, automatically thermostated within ± 0.01 K. Details of the calibration of the apparatus have been reported elsewhere [17]. The repeatability of the $\rho$ measurements is ± 1·10$^{-3}$ kg·m$^{-3}$, while the relative standard uncertainty of the measurements is estimated to be ± 0.12%. The determination of the speed of sound is based on the measurement of the propagation time of short acoustic pulses (3 MHz center frequency [18]), which are repeatedly transmitted to the sample. The repeatability and relative standard uncertainty of the $c$ measurements are respectively, ± 0.1 m·s$^{-1}$ and ±0.0004

The experimental technique was checked by determining $V_m^E$ and $c^E$ of the standard mixture (cyclohexane + benzene) at the temperatures (293.15-303.15) K. Our results agree well with published values [19-21]. The accuracy in $V_m^E$ is less than $\pm(0.012|V_{m,max}^E| + 0.005)$ cm$^3$·mol$^{-1}$, where $|V_{m,max}^E|$ stands for the maximum experimental value of $V_m^E$ with respect to the mole fraction. The $c^E$ accuracy is estimated to be 0.8 m·s$^{-1}$.

Refractive indices were measured using a refractometer model RFM970 from Bellingham-Stanley. The measurement repeatability is ± 0.00004. The measurement method is based on the optical detection of the critical angle at the wavelength of the sodium D line (589.6 nm). The temperature is controlled by Peltier modulus. The temperature stability is ± 0.02 K. The refractometer was calibrated using 2,2,4-trimethylpentane and toluene at the three working temperatures following the recommendations by Marsh [22]. The relative standard uncertainty of the experimental $n_D$ values is ± 0.0015.

### 3. Equations

The thermodynamic properties derived more directly from the experimental measurements obtained by means of the densimeter and sound analyser Anton Paar DSA-5000 are: density, $\rho$, molar volume, $V_m$, coefficient of thermal expansion, $\alpha_p = -\dfrac{1}{\rho}\left(\dfrac{\partial \rho}{\partial T}\right)_p$ and $\kappa_S$. In this work, $\alpha_p$ values were obtained assuming a linear dependence of $\rho$ with $T$. If the absorption of the acoustic wave is negligible, $\kappa_S$ can be calculated from the Newton-Laplace's equation:

$$\kappa_S = \dfrac{1}{\rho c^2} \qquad (1)$$

For an ideal mixture at the same temperature and pressure as the system under study, the values $F^{id}$ of the thermodynamic property, $F$, are calculated using the equations [23-25]:

$$F^{id} = x_1 F_1 + x_2 F_2 \qquad (F = V_m, C_{pm}) \qquad (2)$$

and

$$F^{id} = \phi_1 F_1 + \phi_2 F_2 \qquad (F = \alpha_p; \kappa_T) \qquad (3)$$

where $C_{pm}$ is the isobaric molar heat capacity, $\phi_i = \dfrac{x_i V_{mi}}{V_m^{id}}$ the volume fraction, $\kappa_T$, the isothermal compressibility, and $F_i$, the $F$ value of component i, respectively. For $\kappa_S$ and $c$, the ideal values are calculated according to [23]:

$$\kappa_S^{id} = \kappa_T^{id} - \dfrac{T V_m^{id} \alpha_p^{id 2}}{C_{pm}^{id}} \qquad (4)$$

and

$$c^{id} = \left(\dfrac{1}{\rho^{id} \kappa_S^{id}}\right)^{1/2} \qquad (5)$$

where $\rho^{id} = (x_1 M_1 + x_2 M_2)/V_m^{id}$ ($M_i$, molecular mass of the i component). For $n_D$, ideal values are calculated from the expression [26]:

$$n_D^{id} = [\phi_1 n_{D1}^2 + \phi_2 n_{D2}^2]^{1/2} \qquad (6)$$

The excess functions are determined using the equation:

$$F^E = F - F^{id} \qquad (F = V_m, \kappa_S, c, \alpha_p, n_D) \qquad (7)$$

**4.     Experimental results**

Table 3 lists values, at the working temperatures, of densities, $V_m^E$ and $c$ vs. $x_1$, the mole fraction of DMF (see Figure 1). Table 4 contains the derived quantities $\kappa_S^E$, $\alpha_p^E$ and $c^E$ at 298.15 K (Figures 2-4). Results on $n_D$ and $n_D^E$ are shown in Table 5 (Figure 5). The data were fitted by unweighted least-squares polynomial regression to the equation:

$$F^E = x_1(1-x_1)\sum_{i=0}^{k-1} A_i(2x_1-1)^i \qquad (8)$$

where $F$ stands for the properties cited above. The number of coefficients $k$ used in eq. (8) for each mixture was determined by applying an F-test [27] at the 99.5 % confidence level. Table 6 lists the parameters $A_i$ obtained in the regression and the standard deviations $\sigma$, defined by:

$$\sigma(F^E) = \left[\frac{1}{N-k}\sum(F_{cal}^E - F_{exp}^E)^2\right]^{1/2} \qquad (9)$$

where $N$ is the number of direct experimental values. At $x_1 = 0.5$ and 303.15 K, the $V_m^E$ values reported in the literature, obtained using a dilatometer with Hg as confining liquid, for the systems containing 2-butanone ($-0.3103$ cm$^3$·mol$^{-1}$ [12]) or 2-pentanone ($-0.1802$ cm$^3$·mol$^{-1}$ [13]) are somewhat different to those obtained here: $-0.3345$ and $-0.2630$ cm$^3$·mol$^{-1}$, respectively (see Figure S1 of supplementary material). The $V_m^E$ values derived from $\rho$ measurements for systems with $u = 1, 2$ are very scattered (see Figure S1, supplementary material). No information is given in the original work [11], essentially concerned with conductance measurements, on the experimental technique applied for the density determination.

**5.     Discussion**

Below, we are referring to values of the excess functions at 298.15 K and equimolar composition.

As already mentioned, DMF and ketones are very polar compounds. The dipole moment, $\mu$, of DMF is 3.7 D [28]. The $\mu$ values of 2-alkanones are: 2.69 (acetone); 2.76 (2-butanone); 2.70 (2-pentanone); 2.59 (2-heptanone) D [29]. The impact of polarity on bulk properties is better evaluated by means of the effective dipole moment, $\bar{\mu}$, defined by [30-33]:

$$\bar{\mu} = [\frac{\mu^2 N_A}{4\pi\varepsilon_o V_m k_B T}]^{1/2} \tag{10}$$

where $N_A$, $\varepsilon_o$, $k_B$ stand for the Avogadro's number, the permittivity of the vacuum, and the Boltzmann's constant, respectively. Thus, $\bar{\mu}$ = 1.60 (DMF); 1.2 (propanone); 1.11 (2-butanone); 1.0 (2-pentanone); 0.84 (2-heptanone) (values calculated with data taken from reference [29]). Remarkably, for a given homologous series (2-alkanones) $\mu$ varies only slightly with the chain length of the compound, while the $\bar{\mu}$ variation is much greater. DMF + alkane mixtures show miscibility gaps at rather high temperatures. For example, the upper critical solution temperatures (UCST) of the DMF + heptane, or + hexadecane systems are, respectively, 342.55 K [34], and 385.15 K [35]. Mixtures of acetone with alkanes also show miscibility gaps, but their UCST values are much lower than those containing DMF (245.22 K [36] and 300.9 K [37] for the systems with heptane or hexadecane, respectively). The excess molar enthalpies, $H_m^E$/J·mol$^{-1}$, of 2-alkanone + alkane systems are large and positive: 1704 (propanone) [38]; 1339 (2-butanone) [39]; 886 (2-heptanone) [40]. All these features arise from the existence of strong dipolar interactions between DMF molecules or between ketone molecules in their corresponding mixtures with alkanes. It must be remarked that in systems with a given alkane, say heptane, the $H_m^E$ and $V_m^E$ values decrease with the increasing of $u$. The observed $H_m^E$ variation (see above) [41-43] is due to a weakening of the dipolar interactions between ketone molecules [43]. On the other hand, $V_m^E$ (heptane)/cm$^3$·mol$^{-1}$ = 1.129 (acetone) [44]; 0.280 (2-octanone) [45]. This variation can be ascribed to the positive contribution to $V_m^E$ related to the disruption of ketone-ketone interactions becomes lower when $u$ increases.

DMF + 2-alkanone systems are characterized by low and positive $H_m^E$/J·mol$^{-1}$ values: 37 (acetone) [15], 140 (2-butanone), 206 (2-pentanone) [14]. In view of the features of DMF or 2-alkanone + alkane mixtures, such values reveal the existence of strong amide-ketone interactions. The enthalpy of such interactions may be evaluated from the expression [46-48]:

$$\Delta H_{\text{NCO-CO}} = H_{m1}^{E,\infty}(\text{DMF} + \text{2-alkanone})$$

$$-H_{m1}^{E,\infty}(\text{DMF} + \text{heptane}) - H_{m1}^{E,\infty}(\text{2-alkanone} + \text{heptane}) \tag{11}$$

where $H_{m1}^{E,\infty}$ represents the partial excess molar enthalpy at infinite dilution of the first component. We have applied widely equation (11) for the estimation of the enthalpy of interaction between 1-alkanols and different organic solvents, or, e.g., between amines and ketones [49,50]. Details on its derivation and range of applicability can be found elsewhere [46-48]. For DMF mixtures (see Table 7), $\Delta H_{\text{NCO-CO}}$/kJ·mol$^{-1}$= $-26.0$ (acetone); $-24.1$ (2-butanone) and $-22.6$ (2-heptanone). It is remarkable that these values are more or less similar to those encountered for 1-alkanol + ketone mixtures [47] ($\Delta H_{\text{OH-CO}}$/kJ·mol$^{-1}$= $-28.6$ (methanol + acetone); $-26$ (1-propanol + acetone); $-22.1$ (1-propanol + 2-heptanone). Interestingly, the replacement of DMF by aniline leads to slightly stronger interactions between unlike molecules. Thus, $\Delta H_{\text{N-CO}}$ (aniline)/kJ·mol$^{-1}$= $-29.3$ (propanone); $-24.5$ (2-heptanone) [49]. Amine-ketone interactions are weaker in mixtures where secondary or tertiary amines are involved. For example, $\Delta H_{\text{N-CO}}$ (acetone)/kJ·mol$^{-1}$= $-21.5$ (*N*-methylaniline); $-8.1$ (dipropylamine); $-5.9$ (dibutylamine); $-4.9$ (*N,N,N*-triethylamine) [49,50].

$H_m^E$ values increase with *u*. This is due to the contributions from: (i) the weakening of the amide-ketone interactions (less negative contribution to $H_m^E$); (ii) a larger number of DMF-DMF interactions broken by the longer 2-alkanones (higher positive contribution to $H_m^E$; note that UCST values of DMF + *n*-alkane mixtures increase with the alkane size, see above) are dominant terms over that arising from the disruption of ketone-ketone interactions (lower positive contribution to $H_m^E$).

The excess molar internal energies at constant volume, $U_{Vm}^E$, can be calculated from [33,51]:

$$U_{Vm}^E = H_m^E - \frac{T\alpha_p V_m^E}{\kappa_T} \tag{12}$$

where $\frac{\alpha_p}{\kappa_T} TV_m^E$ is usually termed the equation of state (eos) contribution to $H_m^E$, and $\alpha_p$ and $\kappa_T$ stand for the isobaric thermal expansion coefficient and isothermal compressibility of the mixture, respectively. Along calculations, the needed $\kappa_T$ data were obtained from

$$\kappa_T = \kappa_S + \frac{T\alpha_p^2 V_m}{C_{p,m}} \tag{13}$$

assuming that $C_{pm}^{E} = 0$, which is a good approximation in view of the low $H_m^E$ values of the investigated mixtures [52]. Thus, $U_{Vm}^{E}$/J·mol$^{-1}$ = 207 (acetone); 261 (2-butanone) and 303 (2-pentanone). The large differences between $U_{Vm}^{E}$ and $H_m^E$ values are remarkable, and underline, in the present case, the importance of structural effects on $H_m^E$.

The mixtures have been also investigated by means of the internal pressures, $P_{int}$ [53-56]:

$$P_{int} = \frac{\alpha_p T}{\kappa_T} - p \tag{14}$$

It is known that the main contributions to $P_{int}$ arise from dispersion forces and weak dipole-dipole interactions [55]. For the pure liquids, $P_{int,i}$/MPa = 456.5 (DMF); 331.4 (acetone), 333.4 (2-butanone); 334.5 (2-pentanone) and 326.4 (2-heptanone). The cohesive energy density is defined by [55]:

$$D_{ce} = \frac{\Delta H_{vap} - RT}{V_m} \tag{15}$$

where $\Delta H_{vap}$ is the molar enthalpy of vaporization at 298.15 and $R$ the gas constant. $D_{ce}$ is a measure of the total molecular cohesion (per cm$^3$) of the liquid [55]. Using values of $\Delta H_{vap}$ from [57], $D_{ce,i}$/MPa: 573.8 (DMF), 389.3 (acetone), 358.5 (2-butanone); 334.5 (2-pentanone) and 326.9 (2-heptanone). The comparison between $P_{int,i}$ and $D_{ce,i}$ clearly shows: (i) the existence of very strong dipolar interactions between amide molecules or between ketone molecules for acetone or 2-butanone; (ii) dipolar interactions become weaker when $u$ increases. For the investigated systems, at $x_1 = 0.5$ and 298.15 K, $P_{int}$/MPa = 396.2 (acetone), 388.6 (2-butanone); 386.8 (2-pentanone) and 369.9 (2-heptanone). $P_{int}$ values can be also obtained from the equation [54]:

$$P_{int} = \frac{RT}{x_1 v_{f1} + x_1 v_{f2} + V_m^E} - P \tag{16}$$

In this expression, $v_{fi} (= RT/(p + P_{int,i}))$ is the free volume of component i [54]. From eq. (16), $P_{int}$/MPa = 411.2 (acetone), 404.8 (2-butanone), 401.8 (2-pentanone) and 369.9 (2-

heptanone), which differ by ≈ 4% from those values obtained using eq. (14). This demonstrates the consistency of our data, and that the Van der Waals equation is hold in large extent for the current mixtures, as eq. (16) is derived from this equation of state [54]. We have also determined the excess internal pressures, $P_{\text{int}}^{\text{E}}(= P_{\text{int}} - P_{\text{int}}^{\text{id}})$, with $P_{\text{int}}^{\text{id}} = \dfrac{\alpha_p^{\text{id}} T}{\kappa_T^{\text{id}}} - p$ [58]. Results are shown graphically in Fig. 6. Systems where strong interactions between unlike molecules exist are characterized by large positive $P_{\text{int}}^{\text{E}}$ values. For example, in the case of the aniline + propanone system, $P_{\text{int}}^{\text{E}}$ = 61.35 MPa [59] (Figure 6). Inspection of Figure 6 allows conclude that interactions between unlike molecules become weaker when $u$ increases. The rather low $P_{\text{int}}^{\text{E}}$ value of the 2-heptanone system (7.7 MPa) may be due partially to structural effects as is similar to that for the hexane + hexadecane mixture (6.5 MPa [29,60]) (see below).

Additional information from $P_{\text{int}}$ values can be obtained from the molar refraction, $R_{\text{m}}$, of the systems, which can be calculated using the Lorentz-Lorenz equation [61,62]:

$$R_{\text{m}} = \dfrac{n_{\text{D}}^2 - 1}{n_{\text{D}}^2 + 2} V_{\text{m}} \tag{17}$$

This magnitude is linked to dispersion forces as $n_{\text{D}}$ at optical wavelengths is related to the mean polarizability [61]. As expected, $R_{\text{m}}$, increases with $u$. That is, at such condition, dispersive interactions become more important. On the other hand, the magnitude $P_{\text{int}} V_{\text{m}}$ has been proposed as a measure of the London dispersion energy, independent of the existence of strong specific interactions [63]. In fact, $P_{\text{int}} V_{\text{m}}$ increases linearly with $R_{\text{m}}$ ($P_{\text{int}} V_{\text{m}}$ = 9.15 + 1.15 $R_{\text{m}}$; coefficient of regression, $r$ = 0.997) and decreases linearly with $\bar{\mu}$ ($P_{\text{int}} V_{\text{m}}$ = 65.16-29.58 $\bar{\mu}$; $r$ =0.999). This confirms that dispersive interactions become more relevant when $u$ increases. Therefore, the observed smooth variation of $P_{\text{int}}$ may be due to the increase of dispersive interactions with $u$ is partially counterbalanced by the weakening of the weak dipolar interactions considered within $P_{\text{int}}$.

The parameter $\chi = (\dfrac{c}{c^{\text{id}}})^2 - 1$ is commonly used to estimate the non-ideality of a system. In fact, mixtures which show strong deviations from the ideal behaviour are characterized by high $\chi$ values. Thus, in the case of 1-alkanol + 2-pyrrolidone mixtures, $\chi$ = 0.8 (methanol); 0.35 (ethanol) [64]. For DMF systems, $\chi$ = 0.102 (acetone); 0.072 (2-

butanone); 0.059 (2-pentanone); 0.026 (2-heptanone). In terms of the speed of sound, the present solutions are close to the ideal behaviour.

The Rao's constant, $R$, [65] (or molar sound velocity, $R = V_m c^{1/3}$) is an useful magnitude to investigate molecular interactions in liquid mixtures. This magnitude depends linearly on the molar fractions of the components ($R = x_1 R_1 + x_2 R_2$) if there is no association, or if the degree of association does not change with concentration, as long as the density of the component liquids does not differ largely [65-67]. Systems characterized by complex formation or including a compound with a halogen atom show deviations from this behaviour [68]. Figure 7 shows that $R$ of DMF + 2-alkanone mixtures varies linearly with $x_1$, indicating the absence of complex formation [66,67].

It is well known that negative $V_m^E$ values reveal the existence of interactions between unlike molecules and/or structural effects (geometrical factors as differences in size and shape between them [69-71] or interstitial accommodation [72]). For example, $V_m^E$/cm$^3$·mol$^{-1}$ = $-0.368$ (CHCl$_3$ + 1-butylamine) [73]; $-0.610$ (hexadecane + hexane) [60]. The $H_m^E$/J·mol$^{-1}$ values are $-3125$ ($T$ = 303.15 K) [74] and 114 [75], respectively. The positive $H_m^E$ values and those negative of $V_m^E$ for the investigated systems suggest that the more relevant contribution to this excess function arises from structural effects [71]. Many other mixtures behave similarly. Some examples follow: $H_m^E$/J·mol$^{-1}$ = 322 (*N,N,N*-triethylamine + hexadecane) [76]; 34 (*N,N,N*-tributylamine + octane) [77]; 1780 (2,5,8,11,15-pentaoxapentadecane + pentane) [78]. In the same order, $V_m^E$/cm$^3$·mol$^{-1}$ = $-0.098$ [79]; $-0.070$ [79]; $-0.39$ [78]. We note that the symmetry of the $V_m^E$ curve of the 2-heptanone system is noticeably skewed to higher concentration values of DMF, the smaller component (Figure 1). This is a typical behaviour of systems with compounds which largely differ in size. A similar trend is observed for the systems cited above. Interestingly, both $H_m^E$ and $V_m^E$ of the DMF mixtures increase with $u$. Therefore, the observed $V_m^E$ variation can be ascribed to changes in the interactional contribution to $V_m^E$. This means that the contributions which increase $V_m^E$, weakening of the amide-alkanone interactions and larger number of broken DMF-DMF interactions, are predominant over that arising from the disruption of ketone-ketone interactions, and suggests that the negative $V_m^E$ values are partially due to the rather strong DMF-ketone interactions created upon mixing. We have determined $A_p = (\frac{\Delta V_m^E}{\Delta T})_p = -4.2 \cdot 10^{-3}$; (acetone); $-3.4 \cdot 10^{-3}$ (2-

butanone) and $-2.4 \cdot 10^{-3}$ (2-pentanone) (values in cm$^3$·mol$^{-1}$·K$^{-1}$). Negative $A_p$ and $\alpha_p^E$ values indicate that the structures are more difficult to be broken in the mixtures than in the pure liquids, which may be ascribed to the existence of interactions between unlike molecules. Systems characterized by strong interactions between like molecules show positive values of $A_p$ and $\alpha_p^E$ over the entire mole fraction range. Thus, $A_p$/cm$^3$·mol$^{-1}$·K$^{-1}$ = 7.6·10$^{-3}$ (2-ethoxyethanol + octane) [80] or 2.3·10$^{-3}$ (1-pentanol + cyclohexane) [81]. For the CHCl$_3$ + 1-butylamine mixture, $A_p$ is $-4.2 \cdot 10^{-3}$ cm$^3$·mol$^{-1}$·K$^{-1}$ [73] and is $-1.3 \cdot 10^{-2}$ (same units) for the hexane + hexadecane mixture [60]. The pressure dependence of $H_m^E$ can be evaluated from the expression:

$$\left(\frac{\partial H_m^E}{\partial p}\right)_T = V_m^E - TA_p \tag{18}$$

For the present systems, this magnitude is (in cm$^3$·mol$^{-1}$): 0.821 (acetone); 0.702 (2-butanone) and 0.465 (2-pentanone). The value of the benzene + cyclohexane mixture, characterized by dispersive interactions, is $-0.054$ cm$^3$·mol$^{-1}$ [82]. Mixtures with strong positive deviations from the Raoult's law have more negative values: $-1.54$ (2-ethoxyethanol + octane); $-0.28$ cm$^3$·mol$^{-1}$ (1-pentanol + octane) [80,81]. Values of $\left(\frac{\partial H_m^E}{\partial p}\right)_T$/cm$^3$·mol$^{-1}$ of systems where strong interactions between unlike exist may be rather large and positive: 2.07 (ClCH$_3$ + 1-butylamine [73]); 1.53 (aniline + acetone [83]). Our experimental results seem to support the existence of interactions between unlike molecules, in such way that the increase of pressure leads to the breaking of any type of interactions and $H_m^E$ increases. Negative $\kappa_S^E$ values are also typically ascribed to structural effects and/or to interactions between unlike molecules, while positive values are attributed to the breaking of physical interactions [84]. For example, $\kappa_S^E$/TPa$^{-1}$ = $-142$ (aniline + propanone) [83]; 15.3 (2-ethoxyethanol + octane) [85]. Therefore, the negative $\kappa_S^E$ values of DMF + 2-alkanone systems are in agreement with the trends mentioned above. Finally, we remark the consistency between the signs of the $V_m^E$, $\kappa_S^E$ and $c^E$ functions. For mixtures with acetone, 2-butanone or 2-pentanone, values of $V_m^E$, $\kappa_S^E$ are negative, while those for $c^E$ are positive. In addition, $V_m^E$, $\kappa_S^E$ increase and $c^E$ decreases when $u$ is increased. The behaviour of the 2-heptanone systems is close to the ideal solution.

## 6. Conclusions

Volumetric, speeds of sound and refractive indices data have been reported, at different temperatures, for the systems DMF + acetone, + 2-butanone, + 2-pentanone, or + 2-heptanone. These mixtures are characterized by strong amide-ketone interactions, which become weaker when the alkanone size is increased. The existence of interactions between unlike molecules is supported by positive $P_{int}^E$ values; by negative $V_m^E$ values, as well as by the dependences on temperature and pressure of $V_m^E$. Analysis of the Rao's constant indicates that there is no complex formation. Negative $V_m^E$ values also reveal the existence of structural effects, which have a large contribution to $H_m^E$. $V_m^E$ and $H_m^E$ change in line with the 2-alkanone size. Therefore, the relative $V_m^E$ variation is due to that of the interactional contribution to this excess function. From the $P_{int}V_m$ and $R_m$ values, it is possible to conclude that dispersive interactions become more relevant for the systems with longer 2-alkanones.

TABLE 1

Sample description

| Chemical | CAS number | Source | Purity[a] | Analysis method |
|---|---|---|---|---|
| *N,N*-dimethylformamide | 68-12-2 | Sigma-Aldrich | ≥ 0.995 | GC[b] |
| propanone | 67-64-1 | Sigma-Aldrich | ≥ 0.998 | HPLC[c] |
| 2-butanone | 78-93-3 | Fluka | ≥ 0.995 | GC[b] |
| 2-pentanone | 107-87-9 | Sigma-Aldrich | ≥ 0.98 | FCC[d] |
| 2-heptanone | 110-43-0 | Sigma-Aldrich | ≥ 0.99 | GC[b] |

[a]in mass fraction; [b]gas chromatography; [c] High-performance liquid chromatography
[d]flash column chromatography

TABLE 2

Physical properties[a] of pure compounds at temperature $T$ and pressure $p = 0.1$ MPa

| Property | $T$/K | DMF | propanone | 2-butanone | 2-pentanone | 2-heptanone |
|---|---|---|---|---|---|---|
| $\rho$/g·cm$^3$ | 293.15 | 0.94882 | 0.79019 | 0.80506 | 0.80683 | 0.81539 |
| | | 0.948922[b] | 0.790546[c] | 0.8049[d] | 0.8064[d] | 0.81537[d] |
| | | | 0.78998[d] | 0.80495[e] | 0.80626[e] | 0.815497[f] |
| | | | | 0.805058[f] | | |
| | 298.15 | 0.94406 | 0.78443 | 0.79978 | 0.80199 | 0.81108 |
| | | 0.944163[b] | 0.7844[d] | 0.7997[d] | 0.8015[d] | 0.81123[d] |
| | | | 0.784431[g] | 0.79974[e] | 0.80142[e] | 0.81093[h] |
| | 303.15 | 0.93928 | 0.77863 | 0.79455 | 0.79673 | 0.80681 |
| | | 0.939390[b] | 0.77966[e] | 0.7946[d] | 0.79658[e] | 0.806867[f] |
| | | | | 0.79448[e] | | |
| | | | | 0.794565[f] | | |
| $c$/m·s$^{-1}$ | 293.15 | 1476.7 | 1183.6 | 1212.6 | 1232.4 | 1282.1 |
| | | 1477.8[b] | 1182.5[c] | 1213[e] | 1233[e] | 1281.9[j] |
| | | | 1185[e] | | | |
| | 298.15 | 1458.6 | 1163.6 | 1191.5 | 1211.9 | 1262.7 |
| | | 1458.5[b] | 1161.7[g] | 1192[e] | 1213[e] | 1262.5[j] |
| | | 1458.6[j] | 1163.9[k] | | | |
| | 303.15 | 1438.8 | 1139.9 | 1172.0 | 1191.7 | 1244.4 |
| | | 1439[b] | 1140[e] | 1171[e] | 1192[e] | 1244.1[j] |
| | | 1440.3[j] | 1139.2[c] | | | |
| $\alpha_p$/10$^{-3}$K$^{-1}$ | 298.15 | 1.01 | 1.47;1.45[c] | 1.31 | 1.26 | 1.06; 1.06[d] |
| | | 1.01[j] | 1.46[e] | 1.31[e] | 1.21[e] | |
| $\kappa_S$/TPa$^{-1}$ | 293.15 | 483.34 | 903.4 | 844.8 | 816.1 | 746.1 |
| | | 485[b] | 904.7[c] | 844[e] | 816[e] | 746.3[j] |
| | | | 900[e] | | | |

TABLE 2 (continued)

|  |  |  |  |  |  |  |
|---|---|---|---|---|---|---|
|  | 298.15 | 497.88 | 941.4 | 880.8 | 849.0 | 773.2 |
|  |  | 497.9[b] | 944.6[g] | 880[e] | 848[e] | 773.5[j] |
|  |  | 497.6[j] | 941.1[k] |  |  |  |
|  | 303.15 | 514.28 | 988.4 | 916.3 | 883.8 | 800.4 |
|  |  | 514[b] | 988.8[c] | 918[e] | 883.5[e] | 800.7[j] |
|  |  | 512.9[j] | 987[e] |  |  |  |
| $\kappa_T$/TPa$^{-1}$ | 298.15 | 659.5 | 1323.8 | 1172.8 | 1122.8 | 966.6 |
|  |  | 650[d] | 1324[d] | 1188[d] | 1092[e] | 957[d] |
|  |  | 662[l] |  |  |  |  |
| $C_{p,m}$/ J·mol$^{-1}$·K$^{-1}$ | 298.15 | 146.05[m] | 125.45[n] | 159[o] | 185.4[p] | 242.5[h] |
| $n_D$ | 293.15 | 1.43055 | 1.35845 | 1.37872 | 1.39026 |  |
|  |  | 1.43047[d] | 1.35868[d] | 1.3788[d] | 1.39080[d] |  |
|  |  | 1.4281[q] | 1.3584[r] |  |  |  |
|  | 298.15 | 1.42840 | 1.35386 | 1.37605 | 1.38793 | 1.40688 |
|  |  | 1.4280[l] | 1.35597[s] | 1.3767[t] | 1.3885[t] | 1.40655[d] |
|  | 303.15 | 1.42610 |  | 1.37351 | 1.38523 |  |
|  |  | 1.4271[l] |  | 1.3740[u] | 1.38627[v] |  |

[a] $\rho$, density; $c$, speed of sound; $\alpha_p$, isobaric thermal expansion coefficient; $\kappa_S$, adiabatic compressibility; $\kappa_T$, isothermal compressibility; $C_{p,m}$, isobaric molar heat capacity and $n_D$, refractive index. Relative standard uncertainties, $u_r$, are: $u_r(\rho) = \pm 0.0012$; $u_r(c) = \pm 0.0004$; $u_r(\alpha_p) = \pm 0.028$; $u_r(\kappa_S) = \pm 0.002$; $u_r(\kappa_T) = \pm 0.015$; $u_r(n_D) = \pm 0.0015$; standard uncertainties for temperature and pressure are $u(T) = \pm 0.01$ K (for $n_D$ values, $u(T) = \pm 0.02$ K); and for pressure, $u(p) = \pm 1$ kPa; [b][86] [c] [87]; [d][29]; [e][88]; [f][89]; [g][90] [h][91] [i][17]; [j][92]; [k][93]; [l][94]; [m][95]; [n][96]; [o][97]. [p][98]; [q][99]; [r][100]; [s][101]; [t][102]; [u][103]; [v][104]

TABLE 3

Densities, $\rho$, molar excess volumes, $V_m^E$, and speeds of sound, $c$, for *N,N*-dimethylformamide (1) + 2-alkanone (2) mixtures at temperature $T$ and 0.1 MPa.

| $x_1$ | $\rho$/g·cm$^{-3}$ | $V_m^E$/cm$^3$·mol$^{-1}$ | $c$/m·s$^{-1}$ | $x_1$ | $\rho$/g·cm$^{-3}$ | $V_m^E$/cm$^3$·mol$^{-1}$ | $c$/m·s$^{-1}$ |
|---|---|---|---|---|---|---|---|
| \multicolumn{8}{c}{*N,N*-dimethylformamide (1) + propanone (2) $T$/K= 293.15} ||||||||
| 0.0503 | 0.799565 | −0.0952 | 1200.1 | 0.5476 | 0.883523 | −0.3944 | 1352.9 |
| 0.1057 | 0.809645 | −0.1798 | 1218.0 | 0.5990 | 0.891359 | −0.3696 | 1367.5 |
| 0.1529 | 0.818017 | −0.2349 | 1232.9 | 0.6521 | 0.899454 | −0.3486 | 1382.9 |
| 0.1992 | 0.826191 | −0.2867 | 1247.6 | 0.7051 | 0.907302 | −0.3108 | 1397.8 |
| 0.2475 | 0.834489 | −0.3235 | 1262.6 | 0.7538 | 0.914504 | −0.2803 | 1411.4 |
| 0.2981 | 0.843136 | −0.3600 | 1278.3 | 0.8063 | 0.921993 | −0.2268 | 1425.7 |
| 0.3458 | 0.851059 | −0.3790 | 1292.6 | 0.8441 | 0.927346 | −0.1879 | 1436.0 |
| 0.3942 | 0.859037 | −0.3940 | 1307.4 | 0.9022 | 0.935518 | −0.1273 | 1451.7 |
| 0.4516 | 0.868376 | −0.4059 | 1324.6 | 0.9471 | 0.941723 | −0.0739 | 1463.7 |
| 0.4973 | 0.875647 | −0.4032 | 1338.1 | | | | |
| \multicolumn{8}{c}{*N,N*-dimethylformamide (1) + propanone (2) T/K= 298.15} ||||||||
| 0.0500 | 0.793781 | −0.0952 | 1180.1 | 0.5490 | 0.878710 | −0.4216 | 1336.2 |
| 0.1270 | 0.807814 | −0.2116 | 1205.5 | 0.5893 | 0.884968 | −0.4086 | 1348.1 |
| 0.1485 | 0.811673 | −0.2405 | 1212.5 | 0.6413 | 0.892941 | −0.3849 | 1363.3 |
| 0.1908 | 0.819252 | −0.2963 | 1226.2 | 0.6822 | 0.898928 | −0.3447 | 1374.7 |
| 0.2492 | 0.829349 | −0.3419 | 1244.6 | 0.7435 | 0.908066 | −0.3041 | 1392.3 |
| 0.2970 | 0.837550 | −0.3763 | 1259.6 | 0.7945 | 0.915398 | −0.2507 | 1406.4 |
| 0.3399 | 0.844898 | −0.4091 | 1273.1 | 0.8342 | 0.921109 | −0.2132 | 1417.5 |
| 0.3997 | 0.854838 | −0.4314 | 1291.5 | 0.8801 | 0.927621 | −0.1632 | 1430.1 |
| 0.4461 | 0.862361 | −0.4339 | 1305.3 | 0.9377 | 0.935649 | −0.0932 | 1445.8 |
| 0.4978 | 0.870632 | −0.4308 | 1320.8 | | | | |
| \multicolumn{8}{c}{*N,N*-dimethylformamide (1) + propanone (2) $T$/K= 303.15} ||||||||
| 0.0571 | 0.789325 | −0.1089 | 1159.0 | 0.5422 | 0.872357 | −0.4354 | 1310.9 |
| 0.1075 | 0.798734 | −0.2046 | 1175.8 | 0.6046 | 0.882041 | −0.4074 | 1329.2 |
| 0.1480 | 0.805999 | −0.2542 | 1188.9 | 0.6507 | 0.889133 | −0.3862 | 1342.6 |
| 0.2001 | 0.815316 | −0.3180 | 1205.7 | 0.6986 | 0.896270 | −0.3460 | 1356.3 |
| 0.2528 | 0.824518 | −0.3648 | 1222.4 | 0.7493 | 0.903843 | −0.3080 | 1370.7 |
| 0.3003 | 0.832657 | −0.3953 | 1237.3 | 0.7975 | 0.910848 | −0.2590 | 1384.2 |
| 0.3487 | 0.840866 | −0.4220 | 1252.5 | 0.8442 | 0.917613 | −0.2140 | 1397.1 |
| 0.3920 | 0.848120 | −0.4388 | 1265.8 | 0.8972 | 0.925060 | −0.1448 | 1411.5 |

TABLE 3 (continued)

| | | | | | | | |
|---|---|---|---|---|---|---|---|
| 0.4547 | 0.858369 | −0.4457 | 1284.8 | 0.9411 | 0.931137 | −0.0845 | 1423.4 |
| 0.4991 | 0.865505 | −0.4430 | 1298.1 | | | | |

*N,N*-dimethylformamide (1) + 2-butanone (2) *T*/K= 293.15

| | | | | | | | |
|---|---|---|---|---|---|---|---|
| 0.0482 | 0.811508 | −0.0493 | 1224.5 | 0.5514 | 0.882097 | −0.2945 | 1354.1 |
| 0.1051 | 0.819281 | −0.1116 | 1238.6 | 0.6017 | 0.889386 | −0.2844 | 1367.5 |
| 0.1480 | 0.825263 | −0.1614 | 1249.6 | 0.6519 | 0.896759 | −0.2717 | 1381.2 |
| 0.1915 | 0.831169 | −0.1864 | 1262.8 | 0.7023 | 0.904135 | −0.2460 | 1394.9 |
| 0.2530 | 0.839789 | −0.2364 | 1276.1 | 0.7525 | 0.911634 | −0.2257 | 1408.8 |
| 0.3008 | 0.846349 | −0.2485 | 1288.2 | 0.8002 | 0.918668 | −0.1889 | 1421.7 |
| 0.3386 | 0.851673 | −0.2663 | 1298.0 | 0.8492 | 0.926050 | −0.1546 | 1435.2 |
| 0.3915 | 0.859222 | −0.2904 | 1311.8 | 0.9032 | 0.934175 | −0.1058 | 1450.2 |
| 0.4887 | 0.873095 | −0.3023 | 1337.4 | 0.9477 | 0.940900 | −0.0597 | 1462.4 |

*N,N*-dimethylformamide (1) + 2-butanone (2) *T*/K= 298.15

| | | | | | | | |
|---|---|---|---|---|---|---|---|
| 0.0519 | 0.806811 | −0.0606 | 1204.4 | 0.5536 | 0.877384 | −0.3010 | 1334.6 |
| 0.1048 | 0.814011 | −0.1142 | 1217.5 | 0.5946 | 0.883427 | −0.3002 | 1345.8 |
| 0.1506 | 0.820379 | −0.1637 | 1229.3 | 0.6572 | 0.892579 | −0.2765 | 1362.3 |
| 0.1948 | 0.826510 | −0.2016 | 1240.5 | 0.7034 | 0.899388 | −0.2555 | 1375.5 |
| 0.2482 | 0.834008 | −0.2456 | 1254.4 | 0.7461 | 0.905863 | −0.2432 | 1387.6 |
| 0.3032 | 0.841593 | −0.2619 | 1268.4 | 0.7965 | 0.913261 | −0.1990 | 1401.3 |
| 0.3543 | 0.848769 | −0.2795 | 1281.6 | 0.8488 | 0.921083 | −0.1552 | 1415.8 |
| 0.3962 | 0.854829 | −0.3025 | 1292.8 | 0.8940 | 0.928027 | −0.1235 | 1428.6 |
| 0.4495 | 0.862429 | −0.3095 | 1306.9 | 0.9427 | 0.935405 | −0.0726 | 1442.1 |
| 0.4916 | 0.868442 | −0.3088 | 1318.0 | | | | |

*N,N*-dimethylformamide (1) + 2-butanone (2) T/K= 303.15

| | | | | | | | |
|---|---|---|---|---|---|---|---|
| 0.0496 | 0.801341 | −0.0665 | 1184.6 | 0.5486 | 0.871924 | −0.3341 | 1315.5 |
| 0.1068 | 0.809288 | −0.1410 | 1199.1 | 0.6028 | 0.879882 | −0.3260 | 1330.1 |
| 0.1478 | 0.814925 | −0.1780 | 1209.5 | 0.6517 | 0.886955 | −0.2983 | 1343.2 |
| 0.1836 | 0.819800 | −0.1988 | 1218.5 | 0.7023 | 0.894541 | −0.2826 | 1357.4 |
| 0.2521 | 0.829427 | −0.2550 | 1236.4 | 0.7483 | 0.901328 | −0.2496 | 1369.9 |
| 0.2943 | 0.835351 | −0.2779 | 1248.2 | 0.7995 | 0.908942 | −0.2083 | 1384.1 |
| 0.3532 | 0.843706 | −0.3038 | 1262.8 | 0.8514 | 0.916844 | −0.1727 | 1399.0 |
| 0.3938 | 0.849516 | −0.3199 | 1273.5 | 0.8982 | 0.923832 | −0.1199 | 1410.9 |
| 0.4538 | 0.858110 | −0.3283 | 1289.4 | 0.9460 | 0.931082 | −0.0651 | 1425.2 |
| 0.4969 | 0.864319 | −0.3287 | 1301.1 | | | | |

*N,N*-dimethylformamide (1) + 2-pentanone (2) *T*/K= 293.15

| | | | | | | | |
|---|---|---|---|---|---|---|---|
| 0.1141 | 0.819616 | −0.0909 | 1254.3 | 0.5466 | 0.875212 | −0.2423 | 1350.0 |
| 0.1516 | 0.823961 | −0.1138 | 1261.7 | 0.5987 | 0.882775 | −0.2363 | 1363.1 |
| 0.2063 | 0.830491 | −0.1497 | 1273.0 | 0.6581 | 0.891660 | −0.2236 | 1378.4 |
| 0.2588 | 0.836933 | −0.1821 | 1284.0 | 0.7528 | 0.906494 | −0.1958 | 1404.2 |
| 0.3089 | 0.843115 | −0.1930 | 1294.7 | 0.8029 | 0.914576 | −0.1640 | 1418.1 |
| 0.3594 | 0.849625 | −0.2133 | 1303.2 | 0.8461 | 0.921840 | −0.1434 | 1430.7 |
| 0.4006 | 0.855077 | −0.2283 | 1315.3 | 0.9017 | 0.931277 | −0.0962 | 1446.8 |
| 0.4488 | 0.861489 | −0.2296 | 1326.3 | 0.9496 | 0.939704 | −0.0531 | 1460.9 |
| 0.4999 | 0.868584 | −0.2398 | 1338.5 | | | | |

*N,N*-dimethylformamide (1) + 2-pentanone (2) *T*/K= 298.15

| | | | | | | | |
|---|---|---|---|---|---|---|---|
| 0.0568 | 0.808267 | −0.0472 | 1222.8 | 0.5570 | 0.871899 | −0.2463 | 1332.9 |
| 0.1000 | 0.813221 | −0.0883 | 1231.3 | 0.6022 | 0.878552 | −0.2473 | 1344.4 |
| 0.1526 | 0.819250 | −0.1164 | 1241.7 | 0.6448 | 0.884967 | −0.2436 | 1355.5 |
| 0.1895 | 0.823621 | −0.1408 | 1249.2 | 0.7117 | 0.895208 | −0.2190 | 1373.4 |
| 0.2545 | 0.831549 | −0.1796 | 1262.9 | 0.7504 | 0.901333 | −0.2013 | 1384.0 |
| 0.3298 | 0.841051 | −0.2154 | 1279.3 | 0.8027 | 0.909784 | −0.1704 | 1398.7 |
| 0.3502 | 0.843666 | −0.2208 | 1283.9 | 0.8490 | 0.917570 | −0.1463 | 1412.1 |
| 0.3956 | 0.849557 | −0.2279 | 1294.0 | 0.9019 | 0.926592 | −0.1019 | 1427.6 |
| 0.4567 | 0.857796 | −0.2404 | 1308.5 | 0.9429 | 0.933872 | −0.0708 | 1440.1 |
| 0.5084 | 0.865012 | −0.2502 | 1320.8 | | | | |

*N,N*-dimethylformamide (1) + 2-pentanone (2) T/K= 303.15

| | | | | | | | |
|---|---|---|---|---|---|---|---|
| 0.0581 | 0.803204 | −0.0546 | 1203.0 | 0.5407 | 0.864591 | −0.2649 | 1309.0 |
| 0.1088 | 0.808974 | −0.0954 | 1212.9 | 0.5989 | 0.873078 | −0.2619 | 1323.9 |
| 0.1600 | 0.815007 | −0.1384 | 1223.3 | 0.6467 | 0.880175 | −0.2452 | 1336.3 |
| 0.2044 | 0.820314 | −0.1673 | 1232.5 | 0.6949 | 0.887598 | −0.2312 | 1349.2 |
| 0.2620 | 0.827319 | −0.1923 | 1244.6 | 0.7518 | 0.896616 | −0.2091 | 1364.8 |
| 0.3046 | 0.832689 | −0.2136 | 1253.9 | 0.8007 | 0.904558 | −0.1807 | 1378.6 |
| 0.3506 | 0.838634 | −0.2351 | 1264.1 | 0.8527 | 0.913274 | −0.1469 | 1393.6 |
| 0.4014 | 0.845319 | −0.2496 | 1275.7 | 0.8991 | 0.921307 | −0.1155 | 1407.5 |
| 0.4820 | 0.856310 | −0.2641 | 1294.7 | 0.9485 | 0.929946 | −0.0615 | 1422.2 |
| 0.5006 | 0.858884 | −0.2638 | 1299.1 | | | | |

*N,N*-dimethylformamide (1) + 2-heptanone (2) *T*/K= 298.15

| | | | | | | | |
|---|---|---|---|---|---|---|---|
| 0.0530 | 0.815095 | −0.0062 | 1268.2 | 0.4957 | 0.858130 | −0.0487 | 1328.7 |
| 0.0936 | 0.818298 | −0.0109 | 1272.9 | 0.5985 | 0.871481 | −0.0567 | 1348.2 |

TABLE 3 (continued)

| | | | | | | | |
|---|---|---|---|---|---|---|---|
| 0.1502 | 0.822964 | −0.0158 | 1279.0 | 0.6368 | 0.876894 | −0.0604 | 1356.2 |
| 0.2011 | 0.827389 | −0.0205 | 1285.3 | 0.6999 | 0.886398 | −0.0641 | 1370.3 |
| 0.2495 | 0.831808 | −0.0252 | 1291.3 | 0.7433 | 0.893368 | −0.0630 | 1380.8 |
| 0.2960 | 0.836272 | −0.0309 | 1297.5 | 0.8482 | 0.911958 | −0.0508 | 1408.7 |
| 0.3419 | 0.840896 | −0.0363 | 1304.0 | 0.8634 | 0.914867 | −0.0483 | 1413.1 |
| 0.3963 | 0.846646 | −0.0393 | 1312.2 | 0.8985 | 0.921855 | −0.0412 | 1423.7 |
| 0.4324 | 0.850691 | −0.0457 | 1318.0 | 0.9480 | 0.932305 | −0.0255 | 1439.5 |

[a]standard uncertainties are: $u(x_1) = \pm 0.0008$; $u(p) = \pm 1$ kPa; $u(T) = \pm 0.01$ K; and the relative combined expanded uncertainties (0.95 level of confidence) are: $U_{rc}(\rho) = \pm 0.0024$; $U_{rc}(c) = \pm 0.001$; $U_{rc}(V_m^E) = \pm 0.025$

TABLE 4

Excess functions, at 298.15 K and 0.1 MPa, for $\kappa_S$, adiabatic compressibility, $c$, speed of sound, and $\alpha_p$, isobaric thermal expansion coefficient of N,N-dimethylformamide (1) + 2-alkanone (2) mixtures.

| $x_1$ | $\kappa_S^E$ /TPa$^{-1}$ | $c^E$ /m·s$^{-1}$ | $\alpha_p^E$ /10$^{-3}$·K$^{-1}$ | $x_1$ | $\kappa_S^E$ /TPa$^{-1}$ | $c^E$ /m·s$^{-1}$ | $\alpha_p^E$ /10$^{-3}$·K$^{-1}$ |
|---|---|---|---|---|---|---|---|
| \multicolumn{8}{c}{N,N-dimethylformamide (1) + propanone (2)} ||||||||
| 0.0500 | −16.5 | 9.9 | −0.011 | 0.5490 | −68.5 | 62.9 | −0.047 |
| 0.1270 | −37.6 | 24.1 | −0.025 | 0.5893 | −65.8 | 62.7 | −0.045 |
| 0.1485 | −42.4 | 27.7 | −0.028 | 0.6413 | −61.2 | 61.1 | −0.042 |
| 0.1908 | −50.9 | 34.4 | −0.034 | 0.6822 | −56.3 | 58.4 | −0.038 |
| 0.2492 | −59.4 | 42.2 | −0.040 | 0.7435 | −48.5 | 53.2 | −0.033 |
| 0.2970 | −64.8 | 47.9 | −0.044 | 0.7945 | −40.5 | 46.7 | −0.027 |
| 0.3399 | −68.4 | 52.4 | −0.046 | 0.8342 | −33.8 | 40.5 | −0.023 |
| 0.3997 | −71.2 | 57.4 | −0.048 | 0.8801 | −25.4 | 31.8 | −0.017 |
| 0.4461 | −71.4 | 59.9 | −0.049 | 0.9377 | −13.9 | 18.4 | −0.009 |
| 0.4978 | −70.4 | 61.9 | −0.048 | | | | |
| \multicolumn{8}{c}{N,N-dimethylformamide (1) + 2-butanone (2)} ||||||||
| 0.0519 | −10.2 | 6.7 | −0.007 | 0.5536 | −49.8 | 46.7 | −0.036 |
| 0.1048 | −19.2 | 13.1 | −0.013 | 0.5946 | −48.6 | 47.1 | −0.036 |
| 0.1506 | −26.5 | 18.6 | −0.018 | 0.6572 | −44.8 | 45.7 | −0.033 |
| 0.1948 | −32.3 | 23.3 | −0.023 | 0.7034 | −41.9 | 44.5 | −0.031 |
| 0.2482 | −38.6 | 28.9 | −0.027 | 0.7461 | −38.4 | 42.2 | −0.028 |
| 0.3032 | −43.1 | 33.6 | −0.031 | 0.7965 | −32.6 | 37.6 | −0.024 |
| 0.3543 | −46.4 | 37.5 | −0.033 | 0.8488 | −25.8 | 31.2 | −0.019 |
| 0.3962 | −48.7 | 40.5 | −0.035 | 0.8940 | −19.2 | 24.3 | −0.014 |
| 0.4495 | −50.2 | 43.5 | −0.036 | 0.9427 | −11.0 | 14.6 | −0.008 |
| 0.4916 | −50.5 | 45.2 | −0.037 | | | | |
| \multicolumn{8}{c}{N,N-dimethylformamide (1) + 2-pentanone (2)} ||||||||
| 0.0568 | −7.7 | 5.4 | −0.0021 | 0.5570 | −41.8 | 39.4 | −0.0100 |
| 0.1000 | −13.3 | 9.4 | −0.0039 | 0.6022 | −41.4 | 40.3 | −0.0097 |
| 0.1526 | −19.1 | 14.0 | −0.0058 | 0.6448 | −40.3 | 40.5 | −0.0093 |
| 0.1895 | −22.9 | 17.1 | −0.0068 | 0.7117 | −37.0 | 39.3 | −0.0083 |
| 0.2545 | −29.0 | 22.5 | −0.0083 | 0.7504 | −34.4 | 37.7 | −0.0076 |
| 0.3298 | −34.7 | 28.1 | −0.0095 | 0.8027 | −29.6 | 34.1 | −0.0065 |
| 0.3502 | −35.9 | 29.5 | −0.0097 | 0.8490 | −24.6 | 29.5 | −0.0053 |

TABLE 4 (continued)

| $x_1$ | | | | $x_1$ | | | |
|---|---|---|---|---|---|---|---|
| 0.3956 | −38.1 | 32.2 | −0.0101 | 0.9019 | −17.4 | 22.0 | −0.0037 |
| 0.4567 | −40.7 | 35.8 | −0.0103 | 0.9429 | −10.9 | 14.4 | −0.0023 |
| 0.5084 | −41.6 | 37.9 | −0.0103 | | | | |
| *N,N*-dimethylformamide (1) + 2-heptanone (2) | | | | | | | |
| 0.0530 | −2.4 | 2.0 | | 0.5537 | −18.7 | 18.7 | |
| 0.0936 | −4.4 | 3.7 | | 0.5985 | −19.3 | 19.7 | |
| 0.1502 | −6.4 | 5.4 | | 0.6368 | −19.5 | 20.4 | |
| 0.2011 | −8.6 | 7.4 | | 0.6999 | −19.2 | 20.9 | |
| 0.2495 | −10.4 | 9.0 | | 0.7433 | −18.6 | 20.8 | |
| 0.2960 | −12.0 | 10.7 | | 0.7946 | −17.0 | 19.9 | |
| 0.3419 | −13.6 | 12.3 | | 0.8482 | −14.4 | 17.6 | |
| 0.3963 | −15.3 | 14.1 | | 0.8634 | −13.5 | 16.7 | |
| 0.4324 | −16.3 | 15.3 | | 0.8985 | −11.0 | 14.1 | |
| 0.4957 | −17.7 | 17.2 | | 0.9480 | −6.5 | 8.7 | |

[a]standard uncertainties are: $u(x_1) = \pm 0.0008$; $u(p) = \pm 1$ kPa; $u(T) = \pm 0.01$ K; and the relative combined expanded uncertainties (0.95 level of confidence) are: $U_{rc}(c^E) = \pm 0.04$; $U_{rc}(\kappa_S^E) = \pm 0.05$ and $U_{rc}(\alpha_p^E) = 0.05$

TABLE 5

Excess refractive indices, $n_D$, and the corresponding excess values, $n_D^E$ of N,N-dimethylformamide (1) + 2-alkanone(2) mixtures at temperature $T$ and 0.1 MPa.

| $x_1$ | $n_D$ | $n_D^E$ | $x_1$ | $n_D$ | $n_D^E$ |
|---|---|---|---|---|---|
| \multicolumn{6}{c}{N,N-dimethylformamide (1) + propanone (2) T/K= 293.15} |
| 0.0503 | 1.36286 | 0.00052 | 0.5476 | 1.40156 | 0.00233 |
| 0.1057 | 1.36770 | 0.00112 | 0.5990 | 1.40511 | 0.00222 |
| 0.1529 | 1.37155 | 0.00138 | 0.6521 | 1.40867 | 0.00202 |
| 0.1992 | 1.37532 | 0.00165 | 0.7051 | 1.41226 | 0.00189 |
| 0.2475 | 1.37922 | 0.00193 | 0.7538 | 1.41537 | 0.00161 |
| 0.2981 | 1.38311 | 0.00205 | 0.8063 | 1.41877 | 0.00138 |
| 0.3458 | 1.38685 | 0.00227 | 0.8441 | 1.42113 | 0.00115 |
| 0.3942 | 1.39050 | 0.00236 | 0.9022 | 1.42474 | 0.00079 |
| 0.4516 | 1.39472 | 0.00240 | 0.9471 | 1.42742 | 0.00043 |
| 0.4973 | 1.39800 | 0.00237 | | | |
| \multicolumn{6}{c}{N,N-dimethylformamide (1) + propanone (2) T/K= 298.15} |
| 0.0500 | 1.35901 | 0.00116 | 0.5490 | 1.39951 | 0.00343 |
| 0.1270 | 1.36640 | 0.00247 | 0.5893 | 1.40233 | 0.00328 |
| 0.1485 | 1.36834 | 0.00272 | 0.6413 | 1.40581 | 0.00296 |
| 0.1908 | 1.37223 | 0.00331 | 0.6822 | 1.40851 | 0.00268 |
| 0.2492 | 1.37711 | 0.00367 | 0.7435 | 1.41236 | 0.00210 |
| 0.2970 | 1.38103 | 0.00390 | 0.7945 | 1.41552 | 0.00161 |
| 0.3399 | 1.38441 | 0.00401 | 0.8342 | 1.41805 | 0.00131 |
| 0.3997 | 1.38893 | 0.00400 | 0.8801 | 1.42090 | 0.00092 |
| 0.4461 | 1.39235 | 0.00392 | 0.9377 | 1.42451 | 0.00048 |
| 0.4978 | 1.39595 | 0.00367 | | | |
| \multicolumn{6}{c}{N,N-dimethylformamide (1) + 2-butanone (2) T/K= 293.15} |
| 0.0482 | 1.38120 | 0.00028 | 0.5536 | 1.40712 | 0.00150 |
| 0.1051 | 1.38410 | 0.00055 | 0.6017 | 1.40972 | 0.00146 |
| 0.1480 | 1.38629 | 0.00072 | 0.6519 | 1.41237 | 0.00143 |
| 0.1915 | 1.38843 | 0.00080 | 0.7023 | 1.41495 | 0.00128 |
| 0.2530 | 1.39163 | 0.00105 | 0.7525 | 1.41752 | 0.00110 |
| 0.3008 | 1.39408 | 0.00116 | 0.8002 | 1.42009 | 0.00101 |
| 0.3386 | 1.39604 | 0.00126 | 0.8492 | 1.42274 | 0.00090 |
| 0.3915 | 1.39872 | 0.00131 | 0.9032 | 1.42553 | 0.00060 |
| 0.4432 | 1.40144 | 0.00141 | 0.9477 | 1.42785 | 0.00034 |

TABLE 5 (continued)

| | | | | | |
|---|---|---|---|---|---|
| 0.4887 | 1.40384 | 0.00149 | | | |
| *N,N*-dimethylformamide (1) + 2-butanone (2) *T*/K= 298.15 | | | | | |
| 0.0519 | 1.37886 | 0.00041 | 0.5536 | 1.40522 | 0.00186 |
| 0.1048 | 1.38152 | 0.00059 | 0.5946 | 1.40739 | 0.00185 |
| 0.1506 | 1.38397 | 0.00087 | 0.6572 | 1.41063 | 0.00170 |
| 0.1948 | 1.38631 | 0.00108 | 0.7034 | 1.41301 | 0.00155 |
| 0.2482 | 1.38913 | 0.00132 | 0.7461 | 1.41516 | 0.00133 |
| 0.3032 | 1.39213 | 0.00161 | 0.7965 | 1.41779 | 0.00113 |
| 0.3543 | 1.39479 | 0.00172 | 0.8488 | 1.42048 | 0.00084 |
| 0.3962 | 1.39697 | 0.00178 | 0.8940 | 1.42276 | 0.00050 |
| 0.4495 | 1.39982 | 0.00189 | 0.9427 | 1.42542 | 0.00031 |
| 0.4916 | 1.40199 | 0.00189 | | | |
| *N,N*-dimethylformamide (1) + 2-butanone (2) *T*/K= 303.15 | | | | | |
| 0.0496 | 1.37623 | 0.00079 | 0.5486 | 1.40242 | 0.00419 |
| 0.1068 | 1.37911 | 0.00138 | 0.6028 | 1.40517 | 0.00402 |
| 0.1478 | 1.38113 | 0.00171 | 0.6517 | 1.40780 | 0.00394 |
| 0.1836 | 1.38314 | 0.00221 | 0.7023 | 1.41050 | 0.00373 |
| 0.2521 | 1.38686 | 0.00296 | 0.7483 | 1.41287 | 0.00337 |
| 0.2943 | 1.38911 | 0.00332 | 0.7995 | 1.41553 | 0.00290 |
| 0.3532 | 1.39202 | 0.00352 | 0.8514 | 1.41823 | 0.00231 |
| 0.3938 | 1.39409 | 0.00366 | 0.8982 | 1.42066 | 0.00168 |
| 0.4538 | 1.39728 | 0.00391 | 0.9460 | 1.42320 | 0.00097 |
| 0.4969 | 1.39967 | 0.00413 | | | |
| *N,N*-dimethylformamide (1) + 2-pentanone (2) *T*/K= 293.15 | | | | | |
| 0.0560 | 1.39208 | 0.00015 | 0.5466 | 1.41036 | 0.00123 |
| 0.1141 | 1.39408 | 0.00035 | 0.5976 | 1.41245 | 0.00122 |
| 0.1531 | 1.39541 | 0.00044 | 0.6581 | 1.41495 | 0.00114 |
| 0.2063 | 1.39735 | 0.00066 | 0.6992 | 1.41668 | 0.00106 |
| 0.2588 | 1.39918 | 0.00073 | 0.7528 | 1.41906 | 0.00101 |
| 0.3089 | 1.40106 | 0.00088 | 0.8029 | 1.42129 | 0.00088 |
| 0.3594 | 1.40295 | 0.00097 | 0.8461 | 1.42325 | 0.00074 |
| 0.4006 | 1.40453 | 0.00105 | 0.9017 | 1.42583 | 0.00053 |
| 0.4488 | 1.40643 | 0.00113 | 0.9496 | 1.42811 | 0.00031 |
| 0.4999 | 1.40845 | 0.00118 | | | |

TABLE 5 (continued)

*N,N*-dimethylformamide (1) + 2-pentanone (2) *T*/K= 298.15

| | | | | | |
|---|---|---|---|---|---|
| 0.0568 | 1.38979 | 0.00016 | 0.5570 | 1.40855 | 0.00126 |
| 0.1000 | 1.39128 | 0.00032 | 0.6022 | 1.41039 | 0.00123 |
| 0.1526 | 1.39310 | 0.00047 | 0.6448 | 1.41217 | 0.00119 |
| 0.1895 | 1.39442 | 0.00059 | 0.7117 | 1.41498 | 0.00105 |
| 0.2545 | 1.39670 | 0.00070 | 0.7511 | 1.41672 | 0.00098 |
| 0.3298 | 1.39955 | 0.00092 | 0.8027 | 1.41905 | 0.00088 |
| 0.3502 | 1.40029 | 0.00093 | 0.8490 | 1.42117 | 0.00073 |
| 0.3956 | 1.40203 | 0.00102 | 0.9019 | 1.42364 | 0.00053 |
| 0.4567 | 1.40441 | 0.00110 | 0.9429 | 1.42557 | 0.00032 |
| 0.5084 | 1.40650 | 0.00117 | | | |

*N,N*-dimethylformamide (1) + 2-pentanone (2) *T*/K= 303.15

| | | | | | |
|---|---|---|---|---|---|
| 0.0681 | 1.38748 | 0.00018 | 0.5505 | 1.40580 | 0.00124 |
| 0.1130 | 1.38906 | 0.00035 | 0.5986 | 1.40783 | 0.00126 |
| 0.1576 | 1.39062 | 0.00047 | 0.6407 | 1.40954 | 0.00116 |
| 0.1979 | 1.39212 | 0.00064 | 0.6952 | 1.41187 | 0.00107 |
| 0.2343 | 1.39347 | 0.00076 | 0.7401 | 1.41378 | 0.00091 |
| 0.3128 | 1.39635 | 0.00091 | 0.7992 | 1.41643 | 0.00075 |
| 0.3603 | 1.39820 | 0.00104 | 0.8559 | 1.41908 | 0.00059 |
| 0.4156 | 1.40034 | 0.00112 | 0.8994 | 1.42109 | 0.00037 |
| 0.4485 | 1.40167 | 0.00119 | 0.9411 | 1.42312 | 0.00019 |
| 0.4954 | 1.40355 | 0.00123 | | | |

*N,N*-dimethylformamide (1) + 2-heptanone (2) *T*/K= 298.15

| | | | | | |
|---|---|---|---|---|---|
| 0.0530 | 1.40756 | 0.000034 | 0.5537 | 1.41582 | 0.000193 |
| 0.0936 | 1.40811 | 0.000062 | 0.5985 | 1.41679 | 0.000193 |
| 0.1502 | 1.40890 | 0.000101 | 0.6368 | 1.41766 | 0.000200 |
| 0.2011 | 1.40962 | 0.000114 | 0.6999 | 1.41919 | 0.000207 |
| 0.2495 | 1.41034 | 0.000119 | 0.7433 | 1.42032 | 0.000211 |
| 0.2960 | 1.41107 | 0.000134 | 0.8482 | 1.42329 | 0.000182 |
| 0.3419 | 1.41182 | 0.000143 | 0.8634 | 1.42376 | 0.000177 |
| 0.3963 | 1.41276 | 0.000151 | 0.8985 | 1.42488 | 0.000169 |
| 0.4324 | 1.41341 | 0.000159 | 0.9480 | 1.42651 | 0.000100 |
| 0.4957 | 1.41462 | 0.000171 | | | |

[a] the standard uncertainties, $u$, are: $u(T) = \pm 0.02$ K; $u(p) = \pm 1$ kPa; $u(x_1) = \pm 0.0008$; the relative standard uncertainty of $n_D$ is $u_r(n_D) = 0.0015$; and relative combined expanded uncertainty (0.95 level of confidence) is $U_{rc}(n_D^E) = \pm 0.042$

TABLE 6

Coefficients $A_i$ and standard deviations, $\sigma(F^E)$ (eq. 9) for representation of the $F^{E,a}$ property at 298.15 K and 0.1 MPa for $N,N$-dimethylformamide (1) + 2-alkanone (2) systems by eq. 7

| System[b] | $T$/K | Property $F^E$ | $A_0$ | $A_1$ | $A_2$ | $A_3$ | $A_4$ | $\sigma(F^E)$ |
|---|---|---|---|---|---|---|---|---|
| DMF + propanone | 293.15 | $V_m^E$ | −1.601 | 0.265 | −0.079 | | | 0.003 |
| | | $n_D^E$ | 0.00955 | −0.00144 | | | | 0.00004 |
| | 298.15 | $V_m^E$ | −1.725 | 0.265 | | | | 0.004 |
| | | $c^E$ | 251.1 | 53.9 | | | | 0.6 |
| | | $\kappa_S^E$ | −282.4 | 65.23 | −14.05 | | | 0.17 |
| | | $\alpha_P^E$ | −192.7 | 42.7 | 0.06 | | | 0.00003 |
| | | $n_D^E$ | 0.01492 | −0.00880 | 0.00139 | | | 0.00005 |
| | 303.15 | $V_m^E$ | −1.768 | 0.302 | −0.081 | | | 0.003 |
| DMF + 2-butanone | 293.15 | $V_m^E$ | −1.201 | 0.019 | | | | 0.004 |
| | | $n_D^E$ | 0.00590 | 0.00079 | | | | 0.00004 |
| | 298.15 | $V_m^E$ | −1.246 | 0.021 | | | | 0.005 |
| | | $c^E$ | 181.6 | 69.1 | 26.3 | | | 0.3 |
| | | $\kappa_S^E$ | −202.6 | 3.9 | | | | 0.2 |
| | | $\alpha_p^E$ | −146.8 | −5.3 | −0.05 | | | 0.000001 |
| | | $n_D^E$ | 0.00767 | −0.00017 | −0.00213 | | | 0.00004 |
| | 303.15 | $V_m^E$ | −1.338 | 0.026 | | | | 0.005 |
| | | $n_D^E$ | 0.01644 | 0.00271 | | | | 0.00008 |
| DMF + 2-pentanone | 293.15 | $V_m^E$ | −0.958 | −0.121 | −0.070 | | | 0.003 |
| | | $n_D^E$ | 0.00468 | 0.00152 | | | | 0.00003 |
| | 298.15 | $V_m^E$ | −1.004 | −0.113 | −0.078 | | | 0.003 |
| | | $c^E$ | 150.1 | 78.8 | 39.3 | 17.2 | | 0.1 |
| | | $\kappa_S^E$ | −167.2 | −31 | | | | 0.3 |
| | | $\alpha_p^E$ | −41.3 | −2.4 | −3.2 | | | 0.00011 |
| | | $n_D^E$ | 0.00465 | 0.00155 | | | | 0.00003 |

TABLE 6 (continued)

| | | | | | | | | |
|---|---|---|---|---|---|---|---|---|
| | 303.15 | $V_m^E$ | −1.052 | −0.113 | −0.078 | | | 0.003 |
| | | $n_D^E$ | 0.00498 | −0.00096 | −0.00213 | | | 0.00003 |
| DMF + 2-heptanone | 298.15 | $V_m^E$ | −0.2011 | −0.1822 | −0.1323 | −0.0357 | | 0.0010 |
| | | $c^E$ | 69.4 | 56.4 | 36.8 | 22.5 | 13.2 | 0.1 |
| | | $\kappa_S^E$ | −71.23 | −42.71 | −22.39 | | | 0.11 |
| | | $n_D^E$ | 0.000688 | 0.000373 | 0.000834 | 0.000339 | | 0.000006 |

[a] $F^E = V_m^E$, units: cm$^3$·mol$^{-1}$; $F^E = c^E$, units: m·s$^{-1}$; $F^E = \kappa_S^E$ units: TPa$^{-1}$; $F^E = \alpha_p^E$, units: 10$^{-3}$·K$^{-1}$

TABLE 7

Partial excess molar enthalpies, $H_{m,1}^{E,\infty}$, at T = 298.15 K for solute(1) + organic compound (2) mixtures, and enthalpy of the amide-ketone interaction, $\Delta H_{\text{NCO-CO}}$ (eq. 11), at $T$ = 298.15 K, for amide(1) + 2-alkanone(2) systems.

| System | $H_1^{E,\infty}$ / kJ · mol$^{-1}$ | $\Delta H_{\text{NCO-CO}}$ / kJ·mol$^{-1}$ |
|---|---|---|
| propanone(1) + heptane(2) | 9.09 [90] | |
| 2-butanone(1) + heptane(2) | 7.47 [38] | |
| 2-pentanone(1) + heptane(2) | 6.35 [38] | |
| 2-heptanone(1) + heptane(2) | 5.58 [39] | |
| *N,N*-dimethylformamide(1)+ heptane(2) | 17.1 [105] | |
| *N,N*-dimethylformamide(1) + propanone(2) | 0.2 [14] | − 26.0 |
| *N,N*-dimethylformamide(1) + 2-butanone(2) | 0.5 [14] | − 24.1 |
| *N,N*-dimethylformamide(1) + 2-pentanone(2) | 0.86 [14] | − 22.6 |

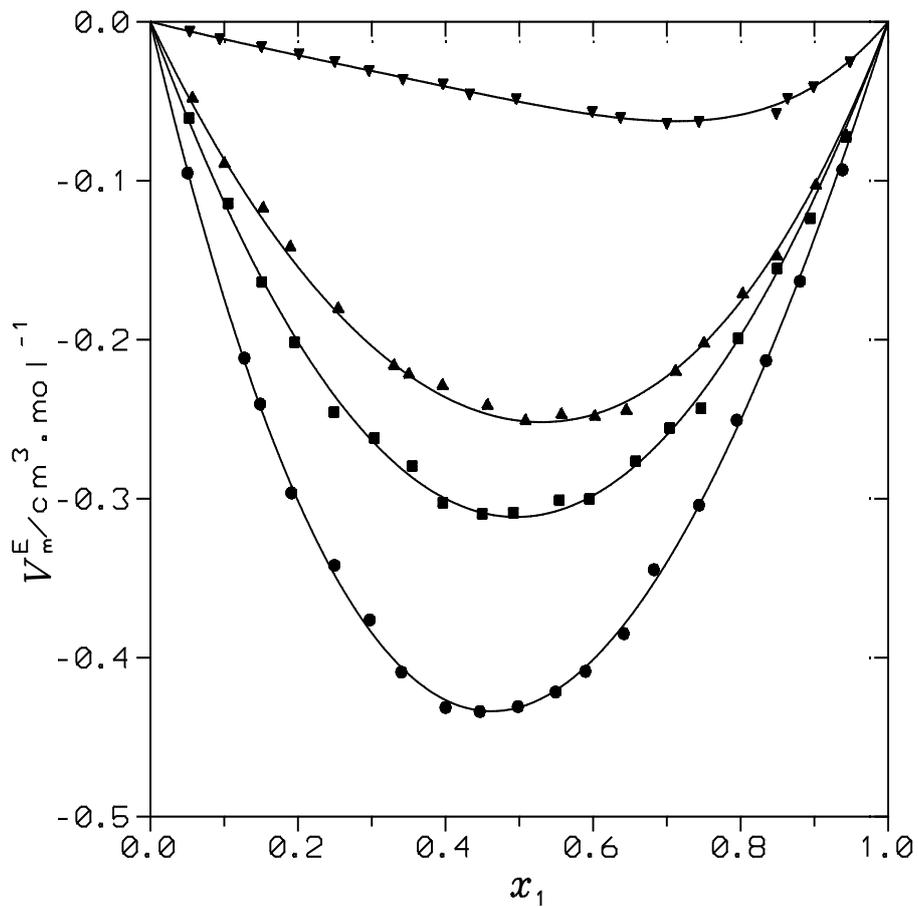

Figure 1    Excess molar volumes, $V_m^E$, for DMF (1) + 2-alkanone (2) systems at atmospheric pressure and 298.15 K. Full symbols, experimental values (this work): (●), propanone; (■), 2-butanone; (▲), 2-pentanone, (▼), 2-heptanone. Solid lines, calculations with eq. 8 using the coefficients from Table 6.

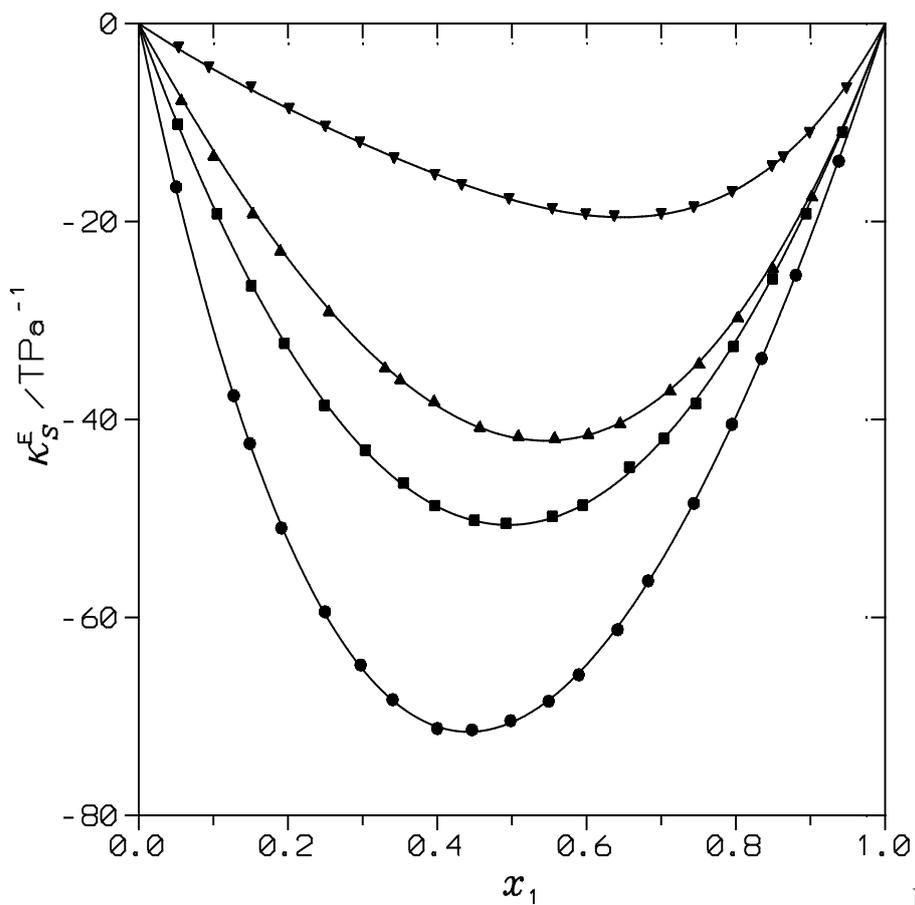

Figure 2 Excess isentropic compressibilities, $\kappa_S^E$, for DMF (1) + 2-alkanone (2) systems at atmospheric pressure and 298.15 K. Full symbols, experimental values (this work): (●), propanone; (■), 2-butanone; (▲), 2-pentanone, (▼), 2-heptanone. Solid lines, calculations with eq. 8 using the coefficients from Table 6.

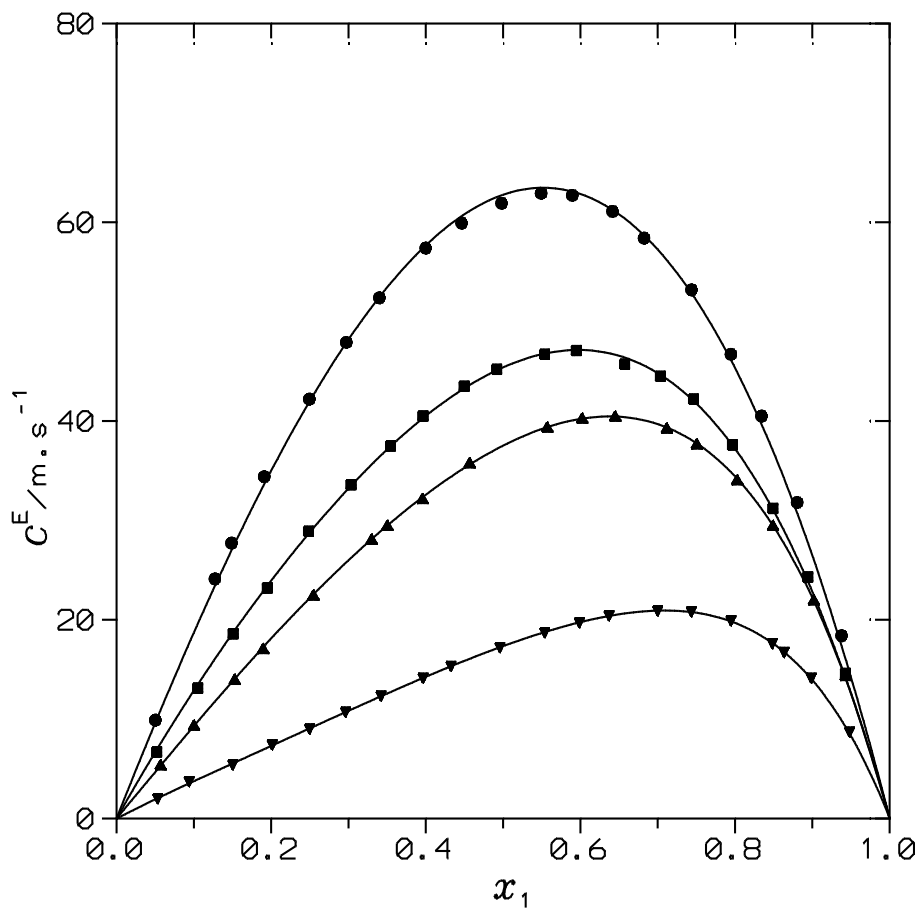

Figure 3    Excess speeds of sound, $c^E$, for DMF (1) + 2-alkanone (2) systems at atmospheric pressure and 298.15 K. Full symbols, experimental values (this work): (●), propanone; (■), 2-butanone; (▲), 2-pentanone, (▼), 2-heptanone. Solid lines, calculations with eq. 8 using the coefficients from Table 6.

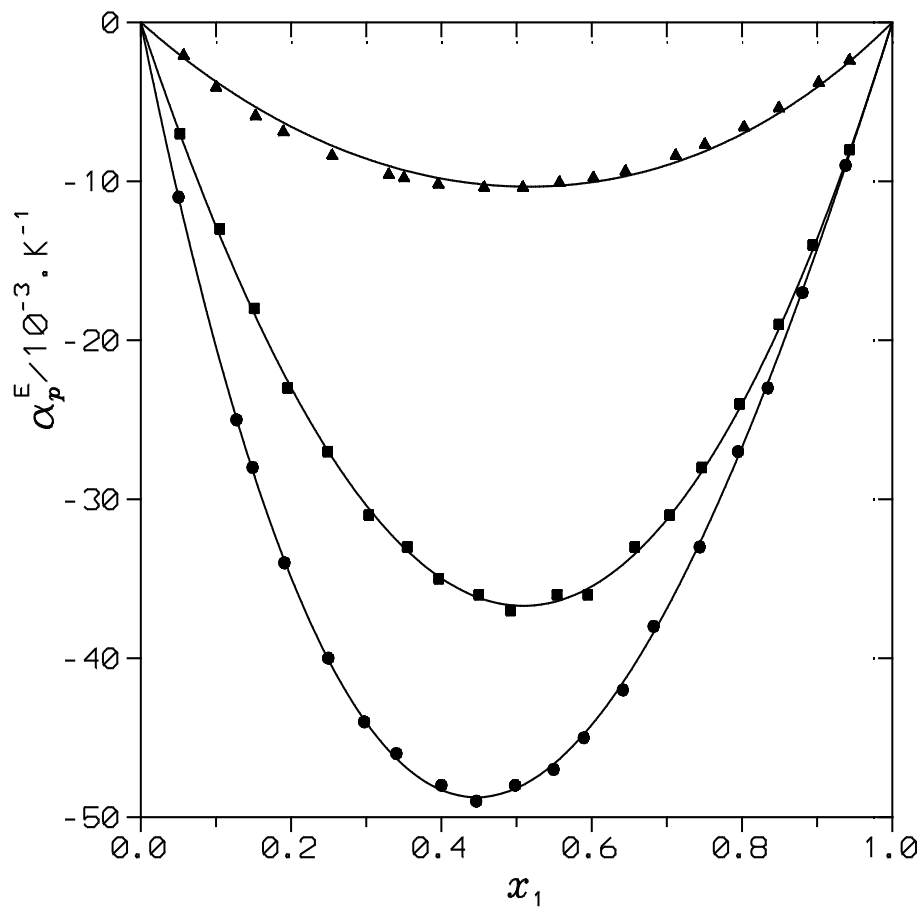

Figure 4　　Excess isobaric thermal expansion coefficients, $\alpha_p^E$, for DMF (1) + 2-alkanone (2) systems at atmospheric pressure and 298.15 K. Full symbols, experimental values (this work): (●), propanone; (■), 2-butanone; (▲), 2-pentanone. Solid lines, calculations with eq. 8 using the coefficients from Table 6.

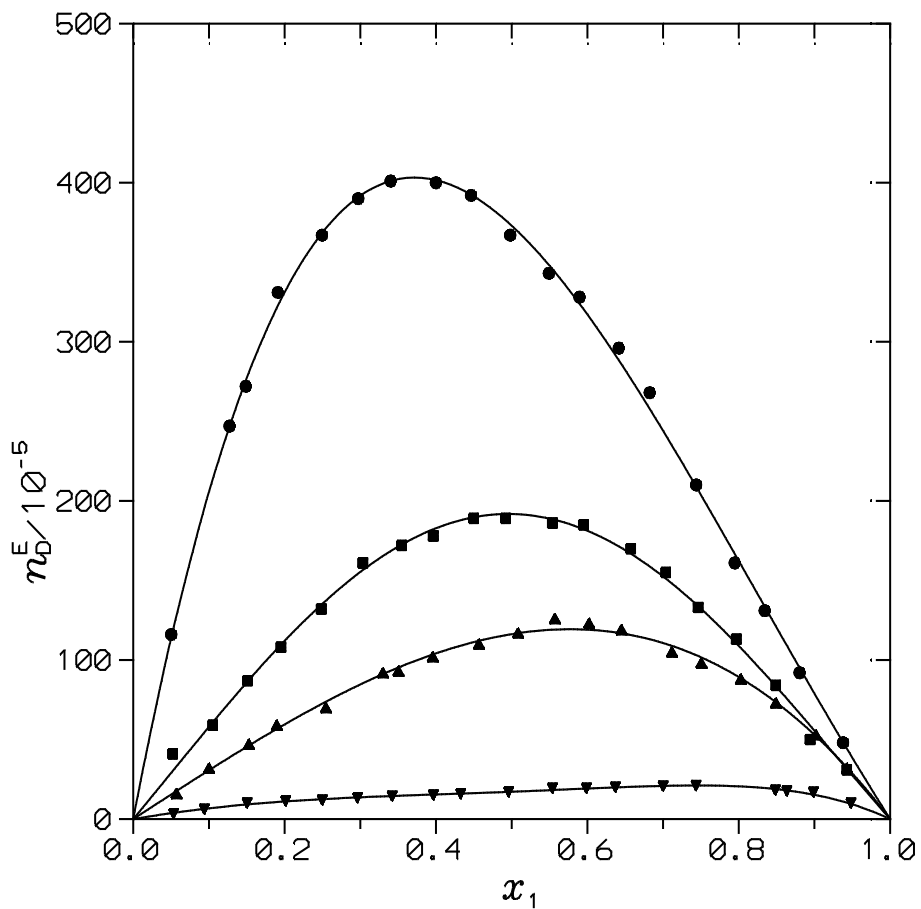

Figure 5    Excess refractive indices, $n_D^E$, for DMF (1) + 2-alkanone (2) systems at atmospheric pressure and 298.15 K. Full symbols, experimental values (this work): (●), propanone; (■), 2-butanone; (▲), 2-pentanone, (▼), 2-heptanone. Solid lines, calculations with eq. 8 using the coefficients from Table 6.

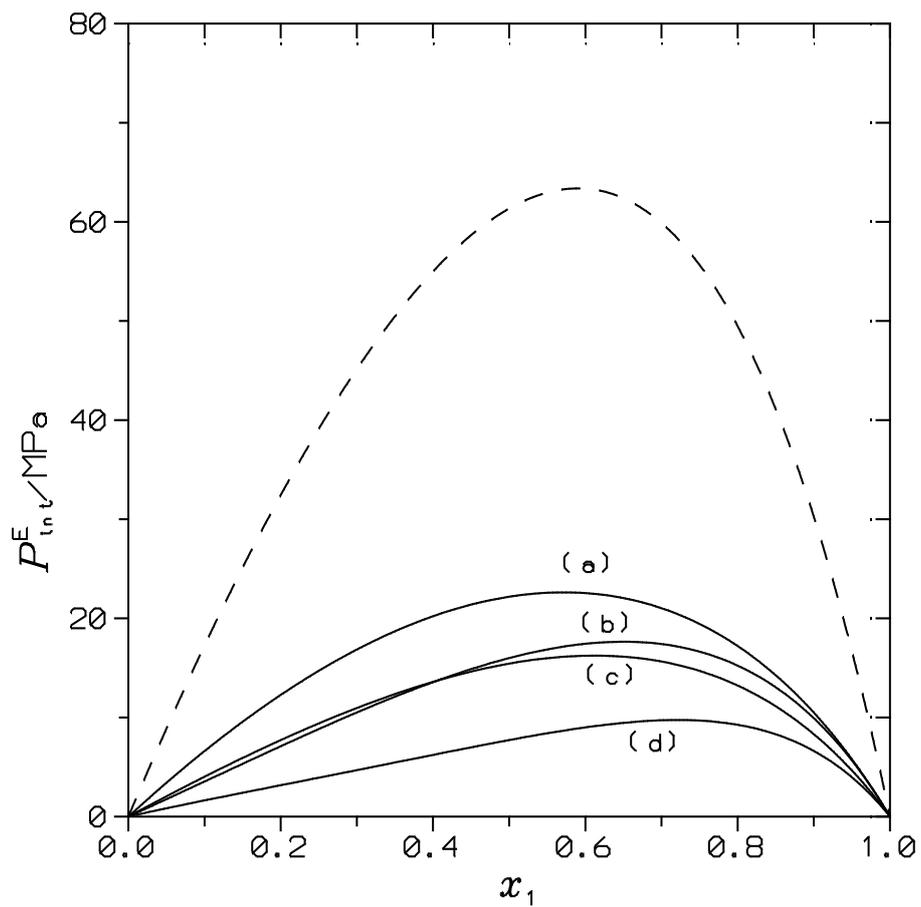

Figure 6    Excess internal pressures, $P_{int}^{E}$, for DMF (1) + 2-alkanone (2) systems at atmospheric pressure and 298.15 K. Solid lines: (a) propanone; (b), 2-butanone; (c), 2-pentanone, (d), 2-heptanone. Dashed line, aniline + propanone [81]

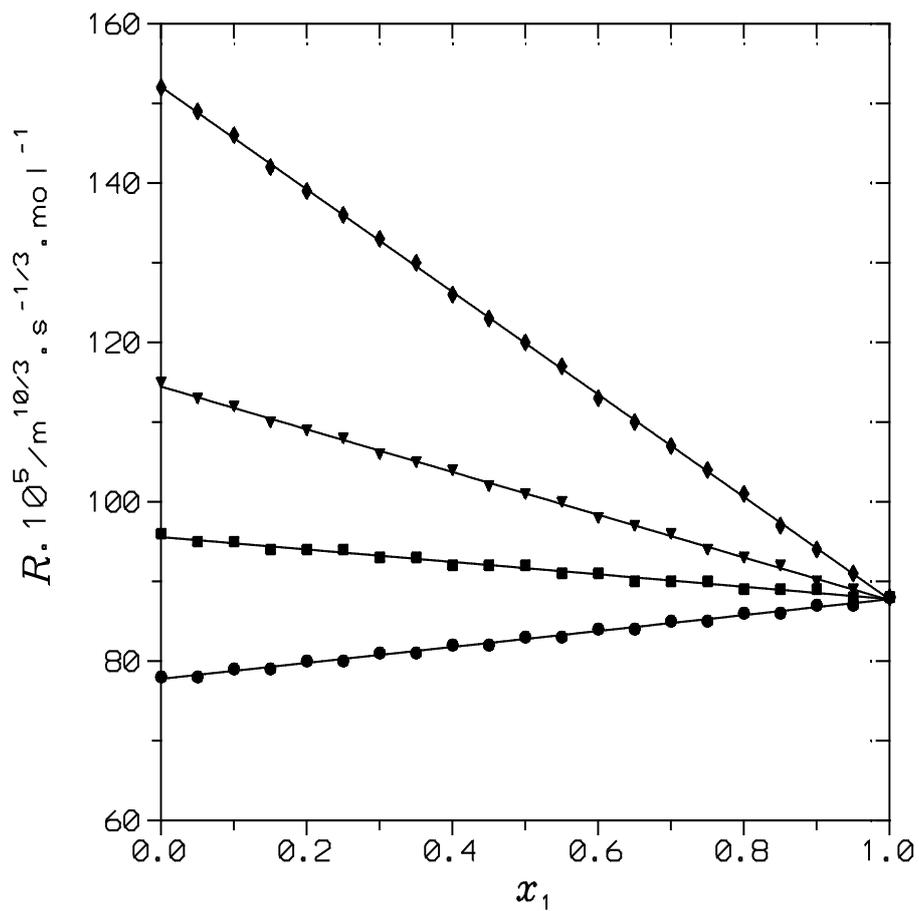

Figure 7    Rao's constant for DMF (1) + 2-alkanone (2) systems at atmospheric pressure and 298.15 K (this work): (●), propanone; (■), 2-butanone; (▲), 2-pentanone, (▼), 2-heptanone.

**SUPPLEMENTARY MATERIAL**

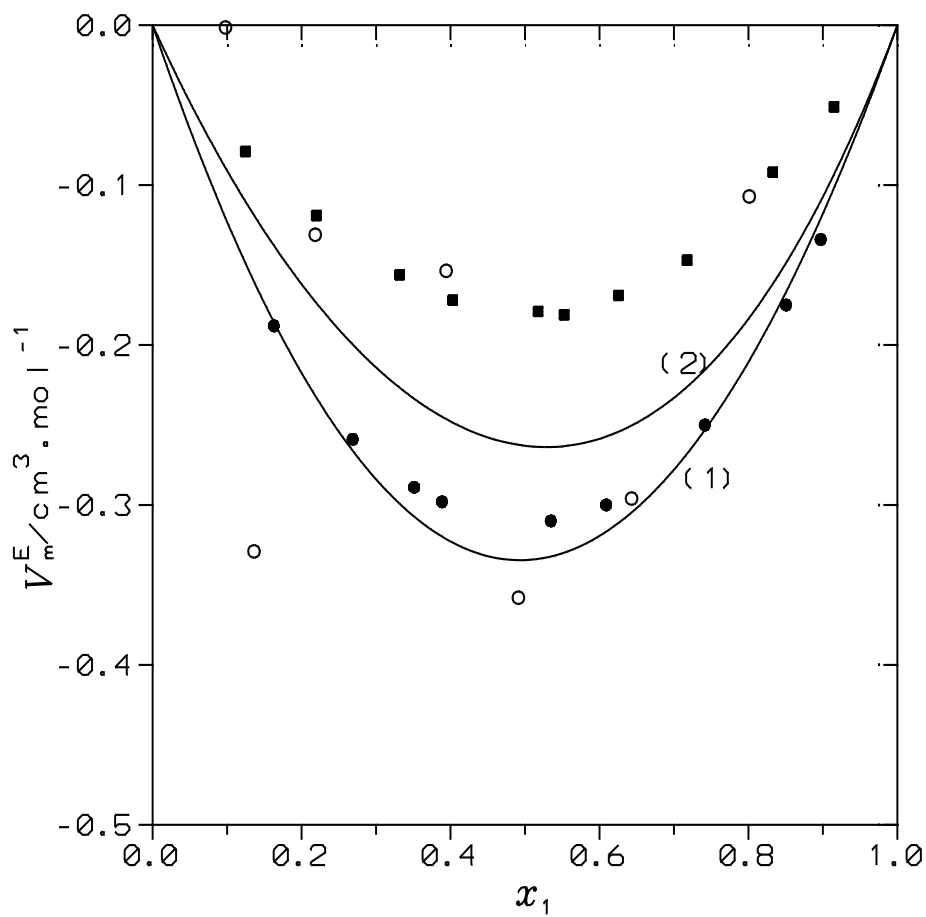

Figure S1   Excess molar volumes, $V_m^E$, for DMF (1) + 2-alkanone (2) systems at atmospheric pressure and temperature $T$. Full symbols, experimental values at 303.15 K: (●), 2-butanone [12]; (■), 2-pentanone [13]. Solid lines, calculations using eq. 8 with coefficients from Table 6: (1), 2-butanone; (2), 2-pentanone. Open symbols: propanone mixture at 298.15 K [11]